\documentclass[sigconf,nonacm=true, natbib=true,anonymous=false]{acmart}

\AtBeginDocument{%
  \providecommand\BibTeX{{%
    \normalfont B\kern-0.5em{\scshape i\kern-0.25em b}\kern-0.8em\TeX}}}

\setcopyright{acmcopyright}
\copyrightyear{2018}
\acmYear{2018}
\acmDOI{10.1145/1122445.1122456}

\acmConference[Woodstock '18]{Woodstock '18: ACM Symposium on Neural
  Gaze Detection}{June 03--05, 2018}{Woodstock, NY}
\acmBooktitle{Woodstock '18: ACM Symposium on Neural Gaze Detection,
  June 03--05, 2018, Woodstock, NY}
\acmPrice{15.00}
\acmISBN{978-1-4503-XXXX-X/18/06}

\usepackage{multirow}
\usepackage[normalem]{ulem}
\usepackage{subcaption}
\usepackage{xcolor}
\usepackage[linesnumbered,ruled]{algorithm2e}
\usepackage{cleveref}
\usepackage{amsmath}
\usepackage{booktabs}
\usepackage{varwidth}
\usepackage{paralist}
\usepackage{bm}

\crefname{equation}{Eq.}{Eq.}
\crefname{section}{Section}{Sections}
\crefname{subsection}{Section}{Sections}
\crefname{subsubsection}{Section}{Sections}
\crefname{figure}{Figure}{Figures}
\crefname{table}{Table}{Tables}
\crefname{subfigure}{Figure}{Figures}
\crefname{algocf}{Algorithm}{Algorithms}

\newcommand{\data}{\textsc{Gest}\xspace}%

\newcommand{\xhdr}[1]{\vspace{0.8mm}\noindent{{\bf #1}}}

\newcommand\blfootnote[1]{%
  \begingroup
  \renewcommand\thefootnote{}\footnote{#1}%
  \addtocounter{footnote}{-1}%
  \endgroup
}

\title{Personalized Showcases: Generating Multi-Modal Explanations for Recommendations}

\author{An Yan$^\ast$, Zhankui He$^\ast$, Jiacheng Li$^\ast$, Tianyang Zhang, Julian McAuley}
\affiliation{\institution{UC San Diego}
  \city{La Jolla}
  \state{CA}
  \country{USA}}
\email{{ayan, zhh004, j9li, tiz010, jmcauley}@ucsd.edu}

\begin{document}

\begin{abstract}

Existing explanation models generate only text for recommendations but still struggle to produce diverse contents.
In this paper, to further enrich explanations, we propose a new task named \emph{personalized showcases}, in which we provide both textual and visual information to explain our recommendations. 
Specifically, we first select a personalized image set that is the most relevant to a user's interest toward a recommended item. 
Then, natural language explanations are generated accordingly given our selected images. 
For this new task, we collect a large-scale dataset from \emph{Google Local} (i.e.,~maps) and construct a high-quality subset for generating multi-modal explanations.
We propose a personalized multi-modal framework which can generate diverse and visually-aligned explanations via contrastive learning.
Experiments show that our framework benefits from different modalities as inputs, and is able to produce more diverse and expressive explanations compared to previous methods on a variety of evaluation metrics.



\end{abstract}
\maketitle
\blfootnote{$^\ast$ equal contribution.}

\section{Introduction}
\begin{figure}[t]
  \includegraphics[width=\linewidth]{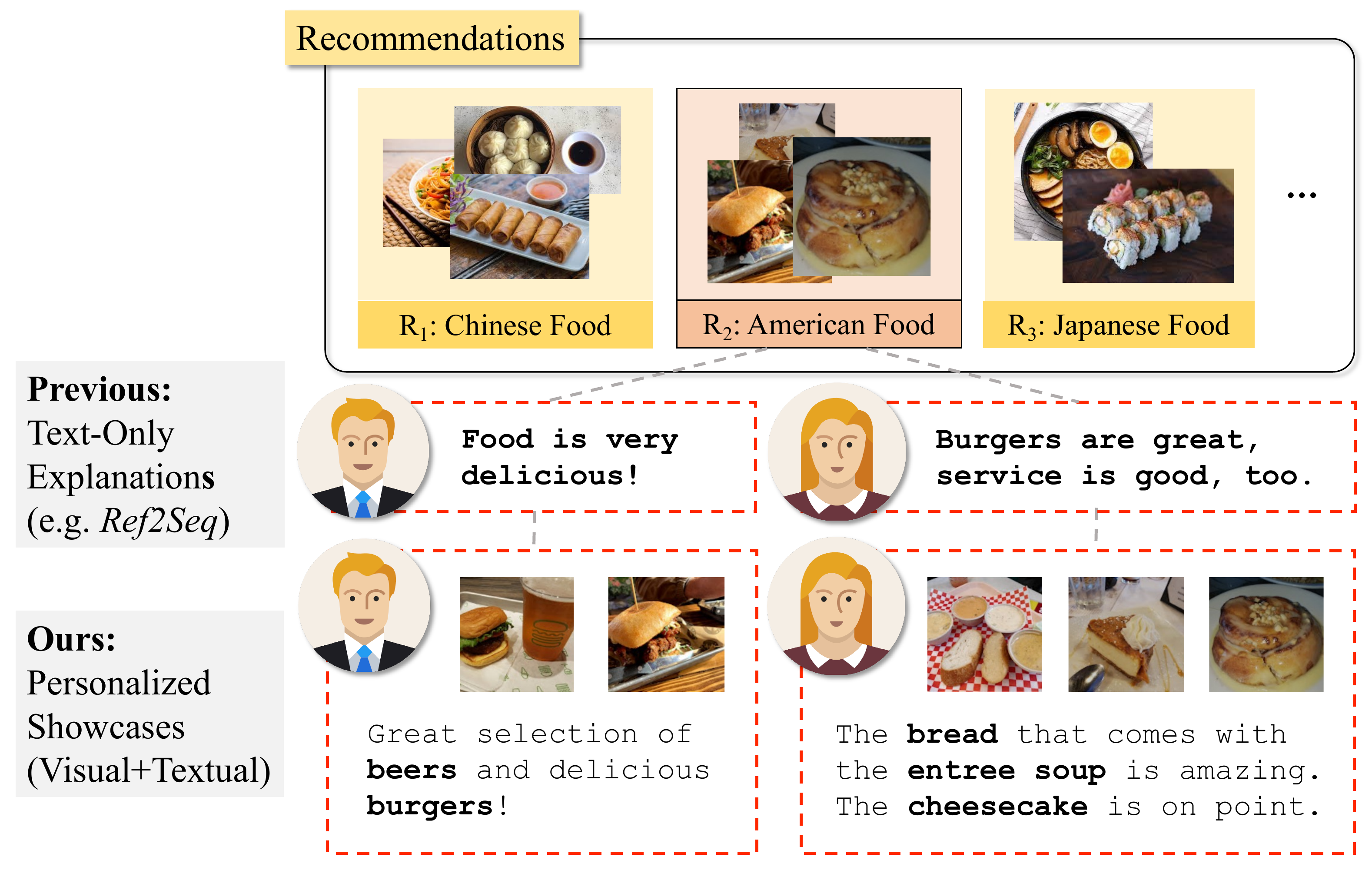}  
  \caption{Illustration of previous text-only explanation and our personalized showcases for recommendations. Given a recommended item or business: (1)~Text-only Explanation models only use historical textual reviews from user and item sides to generate textual explanations. (2)~We propose a personalized showcases task to enrich the personalized explanations with multi-modal (visual and textual) information, which can largely improve the informativeness and diversity of generated explanations.}
  \label{fig:intro}
  \vspace{-1em}
\end{figure}


Personalized explanation generation models have the potential to increase the transparency and reliability of recommendations. Previous works~\citep{Zhou2017LearningTG,Chen2019CoAttentiveML, Baheti2018GeneratingMI, Zang2017TowardsAG} considered generating textual explanations from users' historical reviews, tips~\citep{Li2017NeuralRR} or justifications~\citep{ni2019justifying}. However, these methods still struggle to provide diverse explanations because a large amount of general sentences (e.g.,~`food is very good!') exist in generated explanations and the text generation models lack grounding information (e.g., images) for their generation process.
To further diversify and enrich explanations for recommendations,
we propose a new explanation generation task named \emph{personalized showcases} (shown in~\Cref{fig:intro}). 
In this new task, we explain recommendations via both textual and visual information. 
Our task aims to provide a set of images that are relevant to a user's interest and generate textual explanations accordingly.
Compared to previous works that generate only text as explanations, our showcases present diverse explanations including images and visually-guided text.

To this end, the first challenge of this task is building a
\emph{dataset}. \footnote{Our data is released at \url{https://github.com/zzxslp/Gest} and \url{https://jiachengli1995.github.io/google/index.html}}
Existing review datasets (e.g.,~Amazon~\citep{ni2019justifying} and Yelp
) 
are largely unsuitable for this task~(we further discuss these datasets in~\Cref{sec:data-ana}). Thus, we first construct a large-scale multi-modal dataset, namely \data, which is collected from \underline{G}oogle Local
R\underline{est}aurants including review text and corresponding pictures. 
Then, to improve the quality of \data for personalized showcases, we annotate a small subset to find highly matched image-sentence pairs.
Based on the annotations, we train a classifier with CLIP~\citep{Radford2021LearningTV} to extract visually-aware explanations from the full dataset.
The images and text explanations from users are used as the learning target for personalized showcases. 

\begin{figure*}[t]
    \centering
    \includegraphics[width=\linewidth]{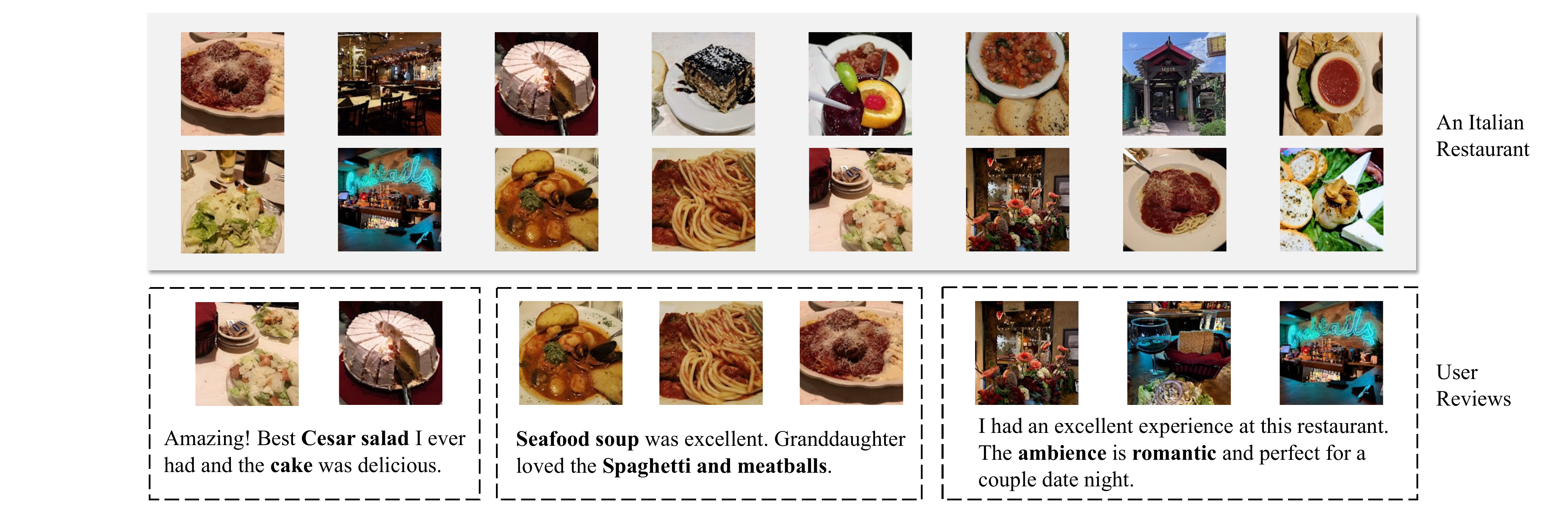}
    \caption{Example of business and user reviews in \data. For a business (e.g.,~an Italian restaurant), \data contains historical reviews and images from different users.}
    \label{fig:reviews}
\end{figure*}

For this new task, we design a new multi-modal explanation framework.
To begin with, the framework selects several images from historical photos of the business that the user is most interested in. Then, the framework takes the displayed images and users' profiles (e.g.,~historical reviews) as inputs and learns to generate textual explanations with a multi-modal decoder. However, generating expressive, diverse and engaging text that will capture users' interest remains a challenging problem. First, 
different from previous textual explanation generation, the alignment between multiple images and generated text becomes an important problem for showcases, which poses higher requirements for information extraction and fusion across modalities.
Second, a typical encoder-decoder model with a cross-entropy loss and teacher forcing can easily lead to generating repetitive and dull sentences that occur frequently in the training corpus (e.g., ``food is great'')~\citep{holtzman2019curious}.

To tackle these challenges, we propose a \textbf{P}ersonalized \textbf{C}ross-Modal \textbf{C}ontrastive \textbf{L}earning ($\mathit{PC^2L}$) framework by contrasting input modalities with output sequences. 
Contrastive learning has drawn attention as a self-supervised representation learning approach~\citep{oord2018representation,chen2020simple}. 
However, simply training with negative samples in a mini-batch is suboptimal~\citep{lee2020contrastive} for many tasks, as the randomly selected embeddings could be easily discriminated in the latent space. 
Hence, we first design a cross-modal contrastive loss to enforce the alignment between images and output explanations, by constructing hard negative samples with randomly replaced entities in the output.
Motivated by the observation that users with similar historical reviews share similar interests, we further design a personalized contrastive loss to reweight the negative samples based on their history similarities.
Experimental results on both automatic and human evaluation show that our model is able to generate more expressive, diverse and visually-aligned explanations compared to a variety of baselines.


Overall, our contributions are as follows:
\begin{itemize}
    \item To generate more informative explanations for recommendations, we present a new task: \emph{personalized showcases} which can provide both textual and visual explanations for recommendations.
    
    \item For this new task, we collect a large-scale multi-modal dataset from \emph{Google Local} (i.e.,~maps). To ensure alignment between images and text, we annotate a small dataset and train a classifier to propagate labels on \data, and construct a high-quality subset for generating textual explanations.
    
    \item We propose a novel multi-modal framework for personalized showcases which applies contrastive learning to improve diversity and visual alignment of generated text. Comprehensive experiments on both automatic and human evaluation indicate that textual explanations from our showcases are more expressive and diverse than existing explanation generation methods. 
\end{itemize}

\section{Task Definition}
\label{sec:task-definition}
In the \emph{personalized showcases} task, we aim to provide both personalized textual and visual explanations to explain recommendations for users. Formally, given user $u\in U$ and business (item) $b\in B$, where $U$ and $B$ are the user set and business set respectively, the \emph{personalized showcases} task will provide textual explanations $S=\{s_1, s_2, ..., s_m\}$ and visual explanations $I = \{i_1, i_2, ..., i_n\}$, where $s$ and $i$ represent sentences and images in explanations. $S$ and $I$ are matched with each other and personalized to explain why $b$ is recommended to $u$.

To better study the relation between textual and visual explanations and provide baselines for future work, in this paper, we decompose the task into two steps as shown in~\cref{fig:model}:
\begin{inparaenum}
\item Selecting an image set as a visual explanation that is relevant to a user's interest;
\item Generating textual explanations given the selected images and a  user's historical reviews.
\end{inparaenum}

Formally, given user $u$, business $b$ and the image candidate set $I_b=\{i^{b}_1, i^{b}_2, \dots i^{b}_{|I_b|}\}$ from $b$, we first select a set of images as visual explanations $I$ from $I_b$ which user $u$ will be interested in, based on user $u$'s profile (i.e.,~historical reviews $X_u=\{x^u_1, x^u_2, ..., x^u_K\}$ and images $I_u = \{i^u_1, i^u_2, ..., i^u_n\}$). 
Then, we use the user's historical reviews $X_u$ and selected images $I$ to generate visually-aware textual explanations $S$.

For our method, we consider the following aspects:
\begin{itemize}
    \item \textbf{Accuracy:} 
    We aim to predict the target images (i.e.,~images associated with the ground-truth review) from business image candidates correctly, and the generated text is expected to be relevant to the business.
    
    \item \textbf{Diversity:}
    The selected images should be diverse and cover more information from businesses (e.g.,~including more dishes from a restaurant). Textual explanations should be diverse and expressive.
    
    \item \textbf{Alignment:}
    Unlike previous explanation or review generation tasks 
    which
    only use historical reviews or aspects as inputs, our visually-aware setting provides grounding to the images. Hence the generated explanations in this new task should aim to accurately describe the content and cover the main objects (e.g.,~the name of dishes, the environment) in the given set of images.
\end{itemize}
\begin{figure}[t]
    \centering
    \includegraphics[width=\linewidth]{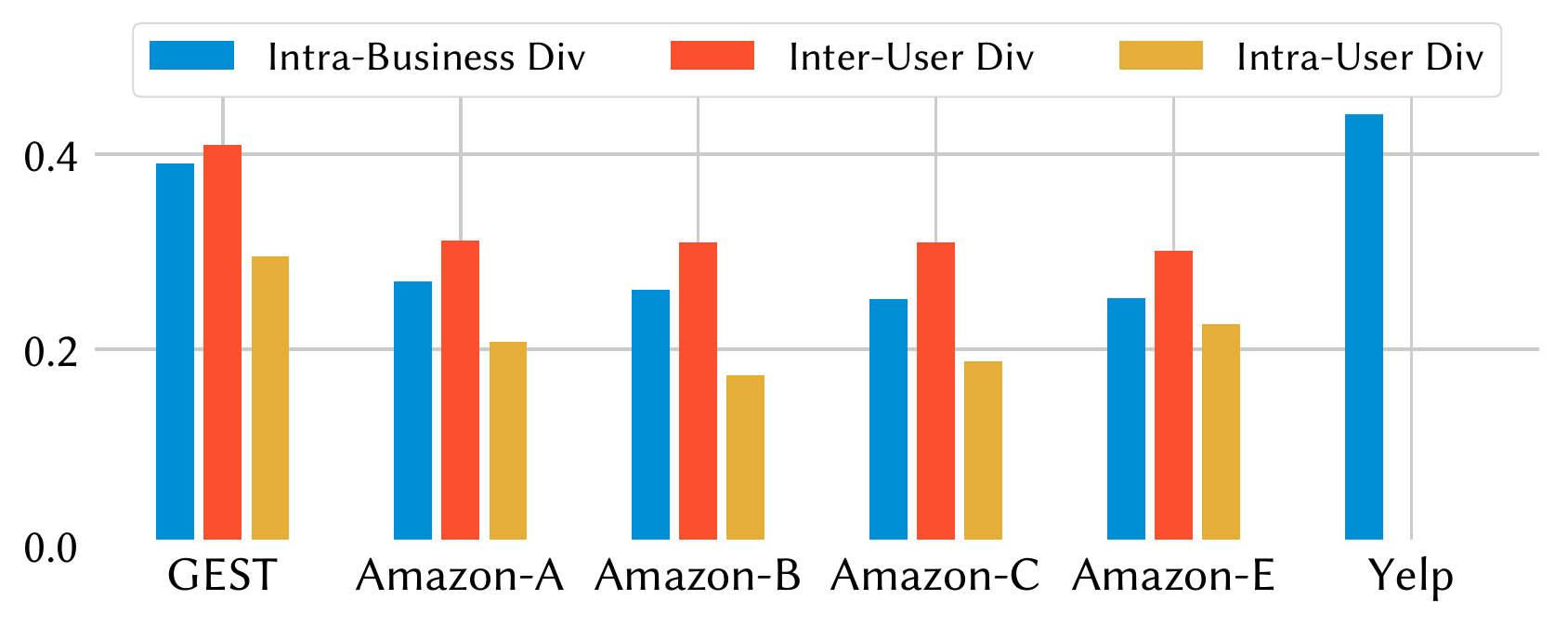}
    \caption{Visual Diversity Comparison. A, B, C, E in Amazon denote different categories of amazon review datasets, which are uniformly sampled from \textit{All}, \textit{Beauty}, \textit{Clothing} and \textit{Electronics}, respectively. Intra-/Inter- User Diversity for the Yelp dataset is unavailable since Yelp images lack user information.}
    \label{fig:data-div}
\end{figure}

\section{Dataset}


\subsection{Dataset Statistics}
\label{sec:data-col}
We collected reviews with images from \emph{Google Local}.
\data-raw in \cref{tab:data} shows the data statistics of our crawled dataset. We can see that \data-raw contains 1,771,160 reviews from 1,010,511 users and 65,113 businesses. Every review has at least one image and the raw dataset has 4,435,565 image urls. 

We processed our dataset into two subsets as (1)~\data-s1 for personalized image set selection, and (2)~\data-s2 for visually-aware explanation generation.
Statistics of our processed dataset are in~\cref{tab:data}, with more processing details in~\Cref{sec:task-2-data} and~\Cref{app:data-collection}.

\subsection{Visual Diversity Analysis}
\label{sec:data-ana}

To distinguish our \data from  existing review datasets and show the usefulness of \emph{personalized showcases}, we first define CLIP-based dis-similarity in three levels to measure the diversity of user-generated images in each business. Then, we compare the visual diversities between our \data data with two representative review datasets, Amazon Reviews~\citep{mcauley2015image,ni2019justifying} and Yelp.

First, similar to~\cite{Radford2021LearningTV,zhu2021imagine}, we use the cosine similarity (denoted as $\textit{sim}$) from pre-trained CLIP to define the dis-similarity between image $i_{m}$ and $i_{n}$ as $\textit{dis}(i_m,i_n)=1-\textit{sim}(i_{m}, i_{n})$.
Thus, we introduce visual diversity in three levels as \emph{Intra-Business Div}, \emph{Inter-User Div} and \emph{Intra-User Div}, which are formally defined in~\Cref{sec:div-def}; higher scores mean more visual diversity.

Then, we investigate the visual diversities for our \data data as well as Amazon Reviews (using all categories \textit{All} (A) and subcategories \textit{Beauty} (B), \textit{Clothing} (C), \textit{Electronics} (E)) and Yelp. For Amazon, we treat each item page as a ``business'' because reviews are collected according to items. In our calculation, we sample 5,000 items with more than one user-uploaded image. Note that images in Yelp dataset do not have user information, so we cannot calculate user-level diversities for Yelp. From~\Cref{fig:data-div}, we have the following observations:

\begin{figure}[t]
    \centering
    \includegraphics[width=0.95\linewidth]{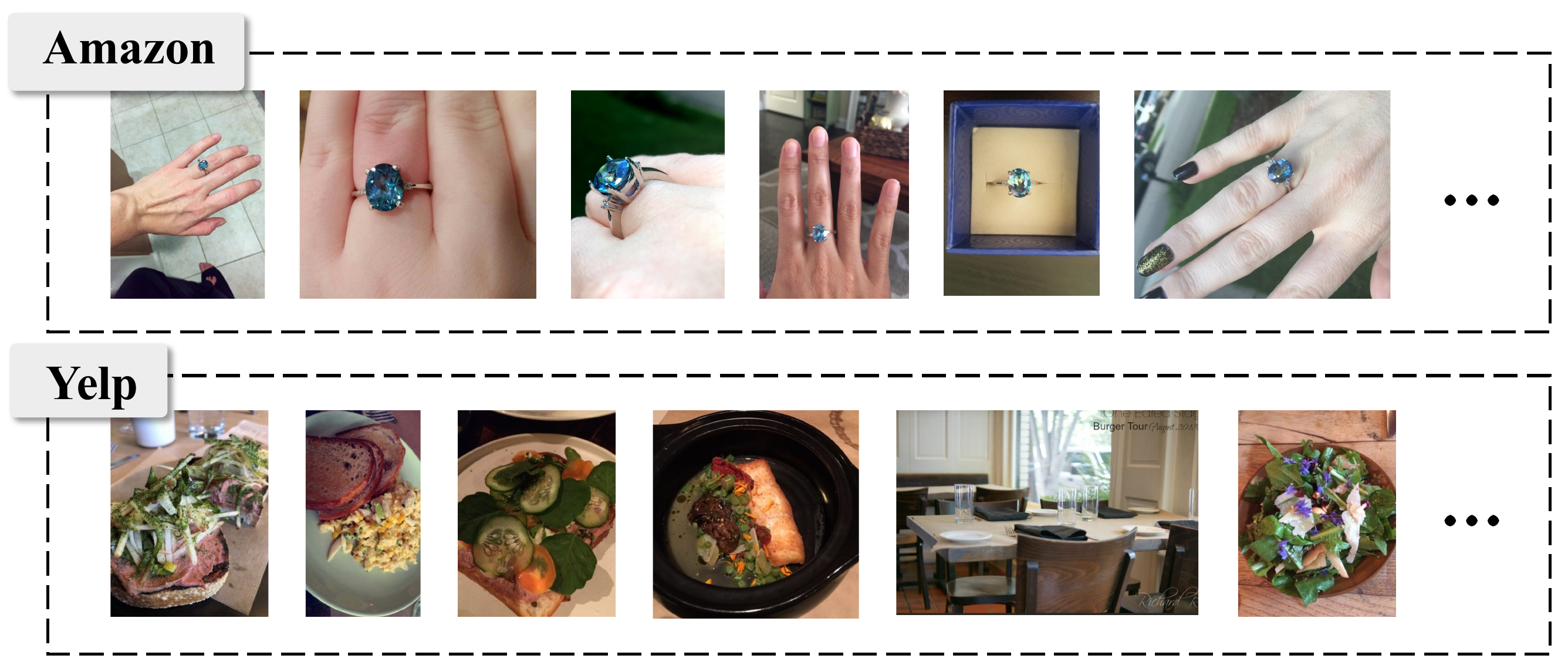}
    \caption{Example of user-generated images from Amazon from an item page and for Yelp from a business. Amazon images mainly focus on a single item and Yelp images for a business are diverse (yet the current public Yelp dataset has no user-image interactions).}
    \label{fig:other_data}
\end{figure}

\begin{table}
\small
\caption{Data statistics for \data. Avg.~R. Len.~denotes average review length and \#Bus.~denotes the number of Businesses. -raw denotes raw \data. -s1 denotes \data data for the first step, and -s2 denotes \data\ data for the second step of our proposed personalized showcases framework. }
\begin{tabular}{lccccc}
\toprule
\textbf{Dataset} & \textbf{\#Image} & \textbf{\#Review} & \textbf{\#User} & \textbf{\#Bus.} & \textbf{Avg. R. Len.} \\ \midrule
\data-raw         & 4,435,565         & 1,771,160         & 1,010,511    &  65,113  & 36.26  \\
\data-s1          & 1,722,296           & 370,563            & 119,086      &  48,330  & 45.48                       \\
\data-s2          & 203,433           & 108,888            & 36,996      &  30,831  & 24.32                       \\ \bottomrule
\end{tabular}
\label{tab:data}
\end{table}

\begin{itemize}
    \item \textbf{Diversities within datasets:} \Cref{fig:data-div} shows that for \data\ and Amazon, \emph{Inter-User Div} is the highest and \emph{Intra-User Div} is the lowest. It indicates even for the same business (item), users focus on and present different visual information. 
    \item \textbf{\data\ \textit{vs.} Amazon:} In~\Cref{fig:data-div}, three visual diversities of Amazon are consistently lower than \data\ by a large margin. We try to explain this by discussing the difference of user behaviors on these two platforms. As an example in~\Cref{fig:other_data}, user-generated images usually focus on the purchased item. Though the information they want to show differs, there is usually a single object in an image (i.e.,~the purchased item). Thus visual diversity is limited. While for \data, as examples in~\Cref{fig:reviews} show, reviews on restaurants allow users to share more diverse information from more varied items, angles or aspects. Compared with Amazon, using \data\ should generate more informative \emph{personalized showcases} according to different user profiles.
    \item \textbf{\data\ \textit{vs.} Yelp:} Yelp images are high-quality (as an example in~\Cref{fig:other_data}) and the \emph{intra-business div.} is higher (0.44) than \data\ (0.39). Images in Yelp themselves are similar to images in \data. However, Yelp images do not fit our task due to the lack of user information. 
\end{itemize}

\begin{figure*}[t]
\centering
\includegraphics[width=0.85\textwidth]{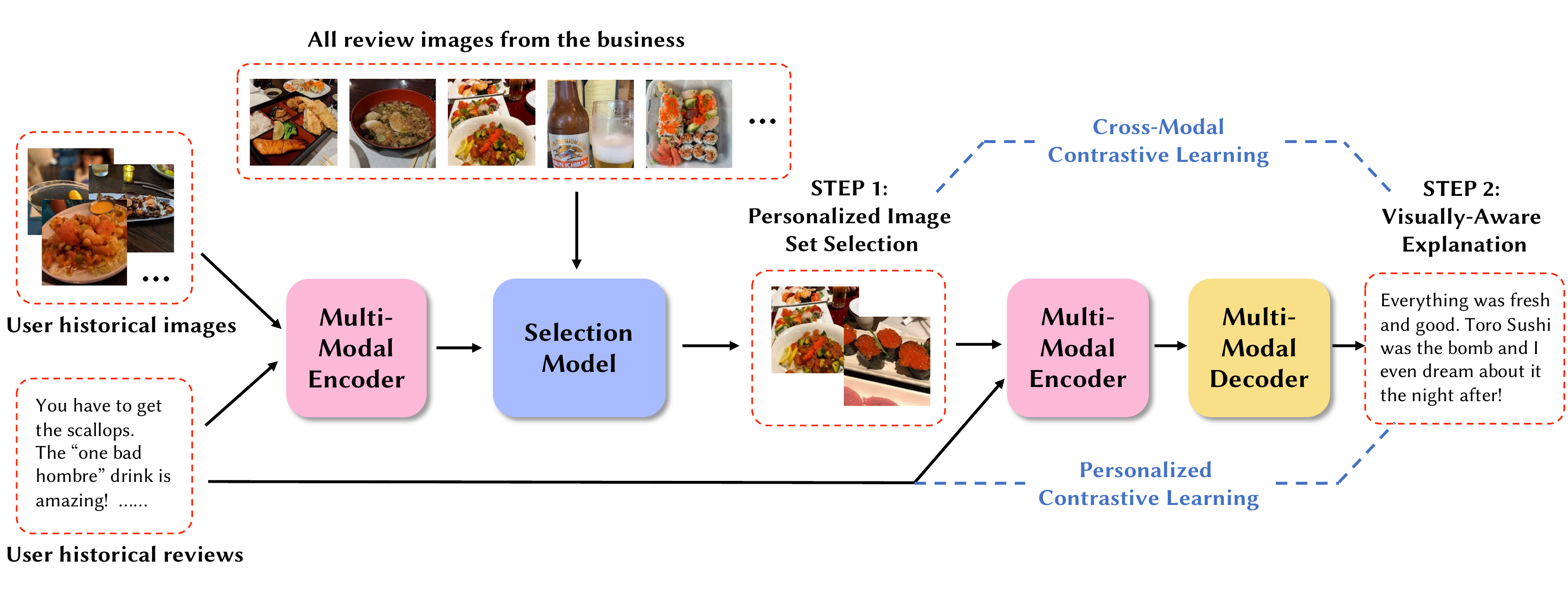}
\caption{Illustration of our \emph{personalized showcases} framework for the given business. We take user historical images and textual reviews as inputs. First, we select an image set that is most relevant to a user’s interest. Then we generate natural language explanations accordingly with a multi-modal decoder. A cross-modal contrastive loss and a personalized contrastive loss are applied between each input modality and the explanations. Last, the selected images and generated textual explanations will be organized as multi-modal explanations to users.}
\label{fig:model}
\end{figure*}

\subsection{Explanation Distillation}
\label{sec:task-2-data}
Reviews often contain uninformative text that is  irrelevant to the images, and cannot be used directly as explanations.
Hence, we construct an explanation dataset from \data-raw. We distill sentences in reviews that align with the content of a given image as valid explanations. 
Three annotators were asked to label 1,000 reviews (with 9,930 image-sentence pairs) randomly sampled from the full dataset. The task is to decide if a sentence describes a image.
Labeling was performed iteratively, followed by feedback and discussion, until the quality was aligned between the three annotators. 
The annotated image-sentence pairs are then split into train, validation, and testing with a ratio of 8:1:1. 

We then train a binary classification model $\Phi$ based on these annotated image-sentence pairs and their corresponding labels. 
Specifically, we extract the embedding of each sentence and image via CLIP. 
The two features are concatenated and fed into a fully connected layer. 
The classifier achieves an AUC of 0.97 and F-1 score of 0.71 on the test set, where similar results are obtained in~\citep{ni2019justifying} for building a text-only explanation dataset.
We use this model to extract explanations from all reviews. The statistics of the dataset \data-s2 can be found in~\cref{tab:data}.

\section{Methodology}
\label{Sec:framework}

In this section, we present our framework of producing personalized showcases. 
As the overview shows~(\cref{fig:model}), we start with personalized image set selection and the visually-aware explanation generation module, then introduce our personalized cross-modal contrastive learning approach in~\cref{sec:contra}.


\subsection{Personalized Image Set Selection}

The first step is to select an image set as a visual explanation that is relevant to a user’s interests,  and is diverse. We formulate this selection step as diverse recommendation with multi-modal inputs.



\xhdr{Multi-Modal Encoder.}
Generally, these user textual- or visual-profiles can be effectively encoded with different pre-trained deep neural networks (e.g.,~ResNet~\cite{he2016deep}, ViT~\cite{Dosovitskiy2021AnII}, BERT~\cite{Devlin2019BERTPO}). Here we choose CLIP~\citep{radford2021learning}, a state-of-the-art pre-trained cross-modal retrieval model as both textual- and visual-encoders. CLIP encodes raw images as image features, and encodes user textual- and visual-profiles as user profile features. 


\xhdr{Image Selection Model.} We use a Determinantal Point Process (DPP) method~\citep{Kulesza2012DeterminantalPP} to select the image subset, which has recently been used for different diverse recommendation tasks~\citep{Wilhelm2018PracticalDR, Bai2019PersonalizedBL}. Compared with other algorithms for \emph{individual} item recommendation, DPP-based models are suitable for \emph{multiple} image selection.  Given user $u$ and business $b$, we predict the image set $\hat{I}_{u,b}$ as follows: 
\begin{equation}
    \hat{I}_{u,b} = \mathrm{DPP}(I_{b}, u),
\end{equation}
where $I_b$ is the image set belonging to business $b$. 
In our design, we calculate user-image relevance using the CLIP-based user's profile features and image features. 
More details of the model are in~\citep{Wilhelm2018PracticalDR}.

\subsection{Visually-Aware Explanation Generation}
\label{sec:step-2}
After obtaining an image set,
we aim to generate personalized explanations given a set of images and a user's historical reviews, with the extracted explanation dataset \data-s2 in~\Cref{sec:task-2-data}.
Specifically, we build a multi-modal encoder-decoder model with GPT-2~\citep{Radford2019LanguageMA} as the backbone.

\xhdr{Multi-Modal Encoder.}
\label{sec:task-2-method}
Given a set of user $u$'s\footnote{We omit the subscript $u$ below for simplicity} historical reviews $X=\{x_1, x_2, \dots, x_K\}$,  we use the text encoder of CLIP to extract the review features $R=\{r_1, r_2, \dots, r_K\}$. 
Similar operations are applied to the input images $I=\{i_1, i_2, \dots, i_n\}$, where we use a pretrained ResNet to extract the visual features $V=\{v_1, v_2, \dots, v_n\}$. Those features are then projected into a latent space: 
\begin{equation}
    Z^{V}_i = W^{V} v_i, Z^{R}_i = W^{R} _i r_i, 
\end{equation}
where $W^{V}$ and $W^{R}$ are two learnable projection matrices. Then we use a multi-modal attention (MMA) module with stacked self-attention layers~\citep{vaswani2017attention} to encode the input features: 
\begin{equation}
    [H^V; H^R] = \mathrm{MMA}([Z^V; Z^R]), 
\end{equation}
where each $H^{V}_i$, $H^{R}_i$ aggregate features from two modalities and $[;]$ denotes concatenation. This flexible design allows for variable lengths of each modality and enables interactions between modalities via co-attentions.

\xhdr{Multi-Modal Decoder.}
Inspired by  recent advances of powerful pre-trained language models, we leverage GPT-2 as the decoder for generating explanations. 
To efficiently adapt the linguistic knowledge from GPT-2, we insert the encoder-decoder attention module into the pre-trained model with a similar architecture in~\citep{Chen2021VisualGPTDA}. 

With this multi-modal GPT-2, given a target explanation $Y=\{y_1, y_2, ..., y_L\}$, the decoding process at each time step $t$ can be formalized as 
\begin{equation}
    \hat{y}_t = \mathrm{Decoder}([H^V; H^R], y_1, \ldots, y_{t-1}).
\end{equation}  

We use a cross-entropy (CE) loss to maximize the conditional log likelihood $\log p_{\theta}(Y|X, I)$ for $N$ training samples ${(X^{(i)}, I^{(i)}, Y^{(i)}})_{i=1}^N$ as follows:
\begin{equation}
    \mathcal{L}_\mathit{CE} = -\sum_{i=1}^N  \log p_\theta (Y^{(i)} | X^{(i)}, I^{(i)}).
\end{equation}
We use ground truth images from the user for training and images from our image-selection model for inference.

\subsection{Personalized Cross-Modal Contrastive Learning}
\label{sec:contra}
Unlike image captioning tasks~\citep{Xu2015ShowAA,yan2021l2c} where the caption is a short description of an image, our task utilizes multiple images as ``prompts'' to express personal feelings and opinions about them. To encourage generating expressive, diverse and visual-aligned explanations, we propose a \textbf{P}ersonalized \textbf{C}ross-Modal \textbf{C}ontrastive \textbf{L}earning ($PC^2L$) framework. We first project the hidden representations of images, historical reviews, and the target sequence into a latent space:
\begin{equation}
    \tilde{H}^V = \phi_V (H^V),\ \tilde{H}^R = \phi_R (H^R),\ \tilde{H}^Y = \phi_Y (H^Y)
\end{equation}
where $\phi_V$, $\phi_R$, and $\phi_Y$ consist of two fully connected layers with ReLU activation~\citep{nair2010rectified} and average pooling over the hidden states $H_V$, $H_R$ and $H_Y$ from the last self-attention layers.
For the vanilla contrastive learning with InfoNCE loss~\citep{oord2018representation, chen2020simple}, we then maximize the similarity between the pair of source modality and target sequence, while minimizing the similarity between the negative pairs as follows:
\begin{equation}
    \mathcal{L}_{\mathit{CL}} = -\sum_{i=1}^N  \log \frac{\exp(s_{i,i}^{X,Y})}{\exp (s_{i,i}^{X,Y}) + \sum\limits_{j\in K} \exp (s_{i,j}^{X,Y})}, 
    \label{eqn:contra}
\end{equation}
where  $s_{i,j}^{X,Y} = \mathit{sim}( \tilde{H}^X_{(i)}, \tilde{H}^Y_{(j)})/\tau$, $\mathit{sim}$ is the cosine similarity between two vectors, $\tau$ is the temperature parameter, $(i)$ and $(j)$ are two samples in the mini-batch, $K$ is the set of negative samples for sample $(i)$.

One challenge of this task is the model is asked to describe multiple objects or contents in a set of images. 
To ensure the visual grounding between multiple image features and output text, we design a novel cross-modal contrastive loss. 
Specifically, given a target explanation $Y=\{y_1, y_2, ..., y_L\}$, we randomly replace the entities~\footnote{We extract entities using spaCy noun chunks~(\url{https://spacy.io/}).} in the text with other entities presented in the dataset to construct a hard negative sample $Y^\mathit{ent}=\{y'_\mathit{ent1}, y_2, ... y'_\mathit{ent2}, ... y_L\}$ (i.e.,~``I like the sushi'' to ``I like the burger''), such that during training, the model is exposed to samples with incorrect entities regarding the images, which are non-trivial to distinguish from the original target sequence. Thus, we add the hidden representation of $Y^\mathit{ent}$ as an additional negative sample $\mathit{ent}$ to formulate the cross-modal contrastive loss:
\begin{equation}
    \mathcal{L}_{\mathit{CCL}} = -\sum_{i=1}^N  \log \frac{\exp(s_{i,i}^{V,Y})}{\exp (s_{i,i}^{V,Y}) + \sum\limits_{j\in K \cup \mathit{ent} } \exp (s_{i,j}^{V,Y})}, 
    \label{eqn:CCL}
\end{equation}
 

On the other hand, to enhance the personalization of explanation generation, we re-weight negative pairs according to user personalities. 
The intuition is that users with more distinct personalities are more likely to generate different explanations. 
Motivated by this, we propose a weighted contrastive loss for personalization:
\begin{equation}
    \mathcal{L}_{\mathit{PCL}} = -\sum_{i=1}^N  \log \frac{\exp(s_{i,i}^{R,Y})}{\exp (s_{i,i}^{R,Y}) + f(i,j) \sum\limits_{j\in K} \exp (s_{i,j}^{R,Y})}.
    \label{eqn:PCL}
\end{equation}
where negative pairs in a mini-batch are re-weighted based on user personality similarity function $f$. In our framework, user personalities are represented by their historical reviews. Specifically, we define $f$ function as: 
\begin{equation}
    f(i,j) = \alpha^{( 1 - \mathit{sim}(\tilde{R}_{(i)}, \tilde{R}_{(j)}))}
\end{equation}
i.e.,~we reduce the weights of negative pairs with similar histories, and increase those with distinct histories. $\alpha$ ($\alpha>1$) is a hyperparameter that weighs the negative samples,  $\mathit{sim}$ is the cosine similarity, $\tilde{R}_{(i)}$ and $\tilde{R}_{(j)}$ are the average features of two users' input historical reviews. 

Overall, the model is optimized with a mixture of a cross-entropy loss and the two contrastive losses:
\begin{equation}
    \begin{split}
    \mathcal{L}_{loss} = \mathcal{L}_{\mathit{CE}} +  \lambda_1 \mathcal{L}_{\mathit{CCL}}  + \lambda_2 \mathcal{L}_{\mathit{PCL}},
    \end{split}
\end{equation}
where $\lambda_1$ and $\lambda_2$ are hyperparameters that weigh the two losses.

\begin{table*}[t]
\centering
\setlength{\tabcolsep}{5pt}
\caption{Results on personalized showcases with different models and different input modalities. Results are reported in percentage~(\%). \textit{GT} is the ground truth.}
\begin{tabular}{l c ccc cc ccc}
\toprule
\multirow{2}{*}{\textbf{Model}} & \multirow{2}{*}{\textbf{Input}} & \multicolumn{3}{c}{\textbf{N-Gram Metrics}} & \multicolumn{2}{c}{\textbf{Diversity Metrics}} & \multicolumn{3}{c}{\textbf{Embedding Metrics}}\\
\cmidrule(lr){3-5} \cmidrule(lr){6-7} \cmidrule(lr){8-10}
  &  & BLEU-1 & METEOR & NIST  & \textsc{Distinct-1} & \textsc{Distinct-2} & \textsc{CLIP-Align}  & \textsc{CLIP-Score} & \textsc{BERT-Score}\\
\midrule
\textit{GT} & - & - &  - & - & 6.06 & 43.23 & 90.47 & 28.41 & - \\
\midrule
\textit{ST} & \textit{img} & 8.24 & 3.41 & 28.08 & 2.74 & 17.41 & 80.84 & 24.31 & 85.20\\
\textit{R2Gen} & \textit{img} & 6.47 & 3.10 & 36.55  & 3.23 & 22.45 & 82.07 & 24.28 & 85.89\\
\midrule
\textit{Ref2Seq} & \textit{text} & 7.09 & 3.80 & 30.78  & 0.92 & 5.89 & 73.51 & 23.83 & 84.71\\
\textit{Peter} & \textit{text} & 8.89 & 3.28 & 34.45  & 0.38 & 1.27 & 72.70 & 23.27 & 86.94\\
\midrule
\multirow{2}{*}{\textit{Ours}}
& \textit{img}  & 9.92 & 3.64 & 37.35 & 3.37 & 26.37 & 84.78 & \textbf{24.68}  & 88.03\\
& \textit{img+text} & \textbf{10.40} & \textbf{3.83} & \textbf{50.64} & \textbf{3.58} & \textbf{28.58}  & \textbf{85.31} & 24.50 & \textbf{88.23}\\
\bottomrule
\end{tabular} 

\label{tab:task-2-main}
\end{table*}

\subsection{A Metric for Visual Grounding}
As mentioned in~\cref{sec:task-definition}, we want our model to generate explanations that can accurately describe the content in a given image set.
Typical n-gram evaluation metrics such as BLEU compute scores based on n-gram co-occurrences, which are originally proposed for diagnostic evaluation of machine translation systems but not capable of evaluating text quality, as they are only sensitive to lexical variation and fail to reward semantic or syntactic variations between predictions and references~\citep{reiter2018structured, zhang2019bertscore, sellam2020bleurt}.
To effectively test the performance of the alignment between visual images and text explanations,  we design an automatic evaluation metric: \textsc{CLIP-Align} based on~\citep{Radford2021LearningTV}.

Given a set of images $I=\{i_1, i_2, ..., i_n\}$ and a set of sentences from the generated text $S=\{s_1, s_2, ..., s_m\}$, we first extract the embeddings of all the images and sentences with CLIP, we compute the metric as follows:
\begin{equation}
    \textsc{CLIP-Align} = \frac{1}{n}\sum_{i=1}^{n} max(\{\mathit{cs}_{1,i}, ..., \mathit{cs}_{m,i}\})
\end{equation}
where $\mathit{cs}_{i,j}$ is the confidence score produced by the CLIP-based classifier $\Phi$ trained on our annotated data. By replacing $\mathit{cs}_{i,j}$ with the cosine similarity of image and sentence embeddings, we obtain another metric \textsc{CLIP-Score}, similar to~\citep{hessel2021clipscore}. 

Compared with previous CLIP-based metrics~\citep{hessel2021clipscore,zhu2021imagine}, \textsc{CLIP-Align} focuses specifically on the accuracy and the alignment between objects in the sentences and the images (e.g. ``food is great'' and ``burger is great'' achieves similar high scores with the same burger image computed on \textsc{CLIP-Score}, and a model that repetitively generates ``food is great'' can reach high performance on \textsc{CLIPscore} in corpus level).
Moreover, the vanilla \textsc{CLIPscore}~\citep{hessel2021clipscore} showed poor correlations with captions containing personal feelings, making it less suitable for this task.
We show in~\cref{sec:experiment} with automatic and human evaluation results that our metric performs better when evaluating alignment between images and text.


\section{Experiments}
\label{sec:experiment}
In this section, we conduct extensive experiments to evaluate the performance of our personalized showcases framework. Ablation studies show the influence of different modalities to personalized showcases. Case studies and human evaluation are conducted to validate that our model present more diverse and accurate explanations than baselines.
\subsection{Experimental Setting}
\label{sec:setup}

\xhdr{Baselines.}
To show the effectiveness of our model, we compare it with a number of popular baselines from different tasks, including image captioning, report generation and explanation generation:
\begin{itemize}
    \item \textit{ST}~\citep{Xu2015ShowAA} is a classic CNN+LSTM model for image captioning.
    \item \textit{R2Gen}~\citep{chen2020generating} is a state-of-the-art memory-driven transformer specialized at generating long text with visual inputs.
    \item \textit{Ref2Seq}~\citep{ni2019justifying} is a popular reference-based seq2seq model for explanation generation in recommendation.
    \item \textit{Peter}~\citep{Li2021PersonalizedTF} is a recent transformer-based explanation generation model which uses the user and item IDs to predict the words in the target explanation.
    \item \textit{img} and \textit{text} refer to image and text features respectively.
\end{itemize}

\xhdr{Evaluation Metrics.} 
For image selection, we report Precision@K, Recall@K and F1@K to measure the ranking quality. Due to the nature of our task, we set a small K ($K=3$).
To evaluate diversity, we introduce the truncated \texttt{div}@K ($K=3$) for the average dissimilarities for all image pairs in recommended images. Formally, given K images $\{i_1, \dots, i_K\}$, \texttt{div}@K is defined as:
    \begin{equation}
        \textit{div}@K = \sum_{1\le m<n \le K}\frac{\textit{dis}(i_m, i_n)}{K(K-1)/2}.
    \end{equation}

For textual explanations, we first evaluate the relevance of generated text and ground truth by n-gram based text evaluation metrics: BLEU~(n=1,4)~\citep{papineni2002bleu}, METEOR~\citep{denkowski2011meteor} and NIST~(n=4)~\citep{doddington2002automatic}. 
To evaluate diversity, we report \textsc{Dinstinct-1} and \textsc{Distinct-2} which is proposed in~\citep{li2015diversity} for text generation models.
We then use CLIP and BERT to compute embedding-based metrics. \textsc{CLIP-Align} is our proposed metrics in~\cref{sec:task-2-method}. \textsc{CLIP-Score}~\citep{hessel2021clipscore} \textsc{BERT-Score}~\citep{zhang2019bertscore} are two recent embedding-based metrics. 

\xhdr{Implementation Details.}
\label{sec:implementation}
We use CLIP~\citep{radford2021learning} with ViT-B/32 as image and text encoder to encode user historical reviews and images. 
We convert user profile feature into a 128-dimensional vector with a MLP model~(1024$\rightarrow$512$\rightarrow$512$\rightarrow$256$\rightarrow$128), and convert candidate images with another MLP~(512$\rightarrow$512$\rightarrow$512$\rightarrow$256$\rightarrow$128), where both models use ReLU activations~\citep{nair2010rectified}. 
We follow~\citep{Wilhelm2018PracticalDR} to calculate each element of $\bm{L}$ and optimize DPP using Adam~\citep{Loshchilov2017FixingWD} with an initial learning rate of 1e-3 and batch size 512. For inference, we use greedy decoding to select $K=3$ images as visual explanation. 

For training $\mathit{PC^2L}$, we use AdamW~\citep{Loshchilov2017FixingWD} as the optimizer with an initial learning rate of 1e-4.
The maximum sequence lengths are set to 64 which covers 95\% of the explanations. 
The maximum number of images and historical reviews are set to 5 and 10 respectively.
The hidden sizes of both the encoder and decoder are 768 with 12 heads. 
There are 3 layers in the encoder and 12 layers in the decoder. 
The batch size for training is 32.
We use the GPT-2-small pre-trained weights with 117M parameters. 
The weighting parameters $\lambda_1$, $\alpha$ and temperature $\tau$ are set to 0.2, 0.2, $e$ and 0.1 respectively.
We use a beam size of 2 for decoding to balance the generation effectiveness and efficiency.

\begin{table}[]
\caption{Ablation study for personalized image selection. Results are reported in percentage~(\%).}
\begin{tabular}{lcccc}
\toprule
                & \multicolumn{3}{c}{\textbf{Accuracy}}                              & \textbf{Diversity}   \\ \cmidrule(l){2-4} \cmidrule(l){5-5} 
\textbf{Method} & Prec@3               & Recall@3             & F1@3                 & Div@3                \\ \midrule
\textit{random}           &   4.87            &     6.14        &    5.43        &      30.24       \\
\midrule
\textit{img}  &      \textbf{25.21}      &     34.05      &   28.97          &     17.12        \\
\textit{text}  &        15.28     &   20.58    &        17.54           &     \textbf{18.68}    \\ 
\textit{img+text}         & \textbf{25.21} & \textbf{34.37} & \textbf{29.09} & 17.07 \\
\bottomrule
\end{tabular}
\label{tab:dpp-ablation}
\end{table}

\subsection{Framework Performance}


We first report the model performance on text evaluation metrics in~\cref{tab:task-2-main}, as we found this last step in our framework came with more challenges and interesting findings, e.g., how to generate human-like explanations and avoid dull text, how to evaluate the generation quality.
Here the input images are selected by our model,\footnote{For effective training and evaluation of our framework, ground truth images of a given user are included in the image candidate pool for selecting. If it is for real-world deployment, ground truth images are not available but similar images can be selected.} 
and the input text consists of historical reviews from users. 

First, the clear gap between text-input models and image-input models on diversity and CLIP-based metrics validates the importance of incorporating image features.
The setting of visually-aware generation models is able to generate accurate explanations with diverse language style.
Second, our \textit{$PC^2L$} shows substantial improvement on most of the metrics compared to LSTM and transformer based models, showing that a pretrained language model with contrastive learning is able to generate high quality explanations.
Finally, though text-based models \textit{Ref2Seq} and \textit{Peter} achieve competitive results with our method on some n-gram metrics such as BLEU and METEOR, their performance is much worse on diversity and embedding metrics. The text quality is also low with repetitive and non-informative sentences appearing often, which we further validate with human evaluations and case studies.


\subsection{Component Analysis}
We conduct ablation studies to evaluate the effectiveness of each component individually.

\xhdr{Model for image set selection.} First, we evaluate the performance of personalized image set selection. For general ranking performance, we compare our model with random selection and different input modalities. As shown in \Cref{tab:dpp-ablation}, though the truncated diversity of the text-only model is the highest, its performance is significantly worse than those with images in terms of ranking metrics. This indicates text input alone is far insufficient to provide personalization for users, and its recommendation result is closer to that of random selection. Historical images on the other hand, provide an important visual cue for modeling users' preference. Overall, a model with images and text can achieve the best ranking performance for image set selection, which validates the importance of our multi-modal setting for personalized showcases.

\begin{table}[]
\caption{Ablation study on contrastive learning. Baseline is to train a multi-modal decoder without contrastive learning. CL, CCL and PCL are the contrastive losses in~\cref{eqn:contra},~\cref{eqn:CCL} and~\cref{eqn:PCL}}
\begin{tabular}{lccccc}
\toprule
\textbf{Method}  & BLEU-1 & \textsc{Distinct-2} & \textsc{CLIP-Align} \\
\midrule
\textit{Baseline}   &  7.96 &  25.90   &  82.50  \\
\midrule
\textit{img CL + text CL}  & 9.72  & 27.58 &  84.03\\
\textit{CCL+ text CL}      & 10.19  & 28.10 & 85.12  \\
\textit{img CL + PCL}      & 9.96  & 28.32  & 84.15 \\
\midrule
\textit{$PC^2L$}     & \textbf{10.40}  &  \textbf{28.58} & \textbf{85.31} \\
\bottomrule
\end{tabular}
\label{tab:contra}.
\end{table}

\begin{figure}[t]
    \centering
    \includegraphics[width=\linewidth]{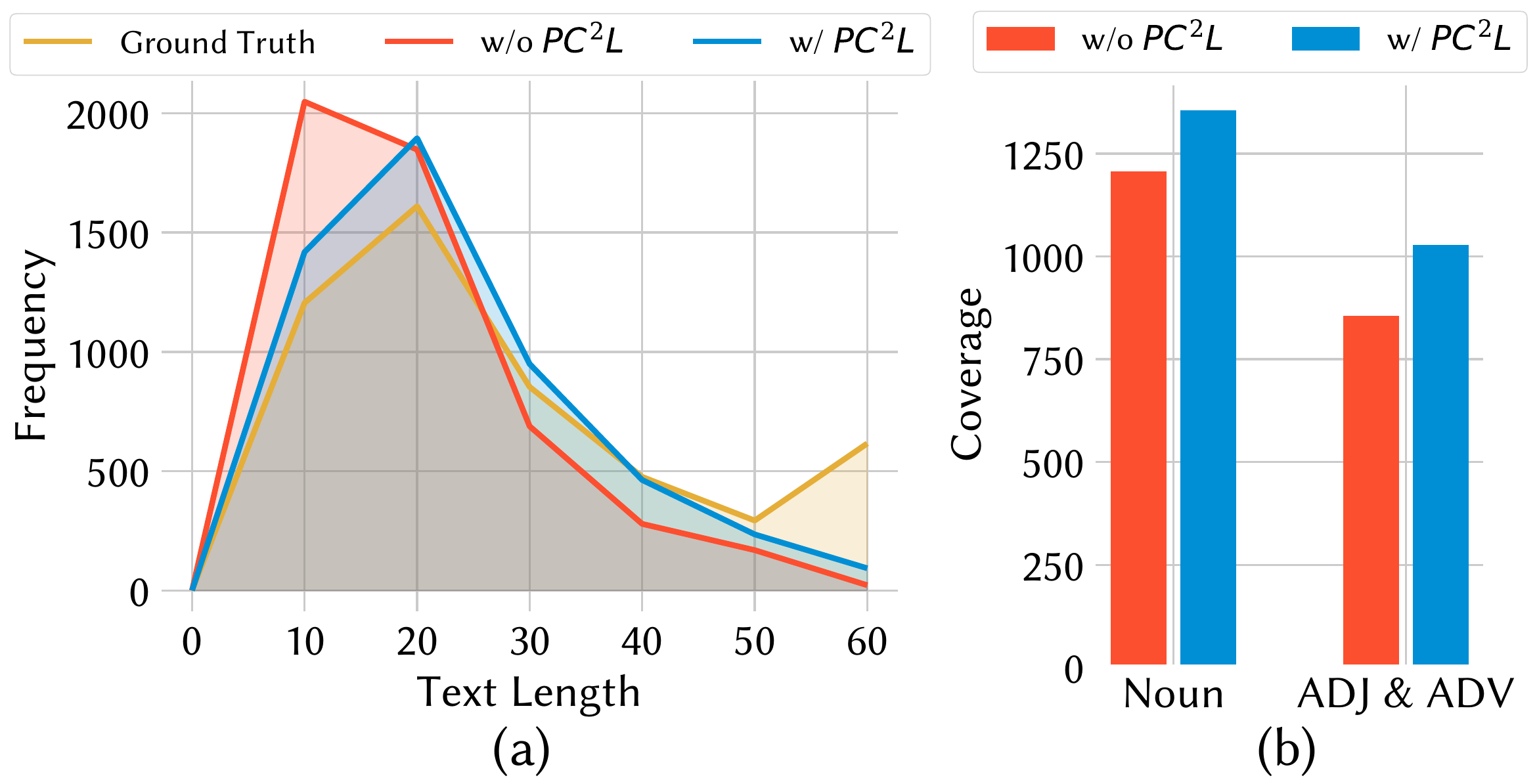}
    \caption{(a) The length distributions of generated texts on the test set. (b) The generated explanation coverage of nouns (Noun), adjectives (ADJ) and adverbs (ADV) in ground truth.}
    \label{fig:dist}
    \vspace{-1em}
\end{figure}

\begin{figure*}[t]
\vspace{-3ex}
\centering
\includegraphics[width=\textwidth]{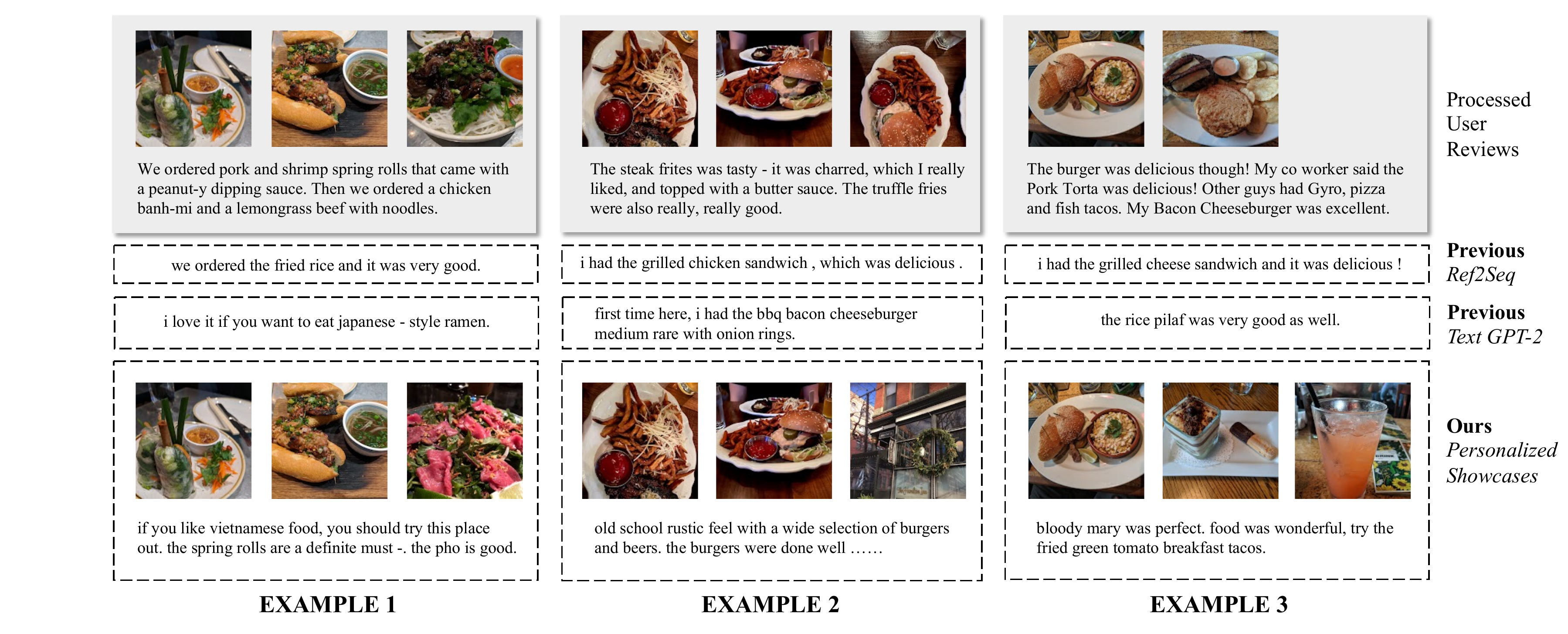}
\caption{Comparison between text-only explanations (i.e.,~\textit{Ref2Seq} and \textit{Text GPT-2}) and our showcases. User reviews are processed following~\Cref{sec:task-2-data}.}
\label{fig:case_study}
\end{figure*}

\xhdr{Effectiveness of Contrastive Learning}
We conduct ablation studies on different variations of our contrastive loss to verify the effectiveness of our method.
As shown in~\cref{tab:contra}, our $\mathit{PC^2L}$ achieves the best performance over all baselines on different metrics.
Specifically, \textit{CCL} contributes more to the visual grounding by enforcing the model to distinguish random entities from the correct ones, and improves \textsc{CLIP-Align} compared to the vanilla contrastive framework~\citep{chen2020simple}. \textit{PCL} improves more on diversity by encouraging the model to focus on users with dissimilar interest.

To further evaluate the generation quality improved by contrastive learning, we analyze the generated explanations from two aspects, length distributions of generations and keywords coverage. \Cref{fig:dist} (a) compares the length distributions of generations on the test set to the ground truth. 
We categorize text lengths into 6 groups (within the range of [0, 60] with an interval of 10). 
The model without $\mathit{PC^2L}$ has a sharper distribution, while adding our $\mathit{PC^2L}$ leads to a distribution which is closer to the ground truth, demonstrating its effectiveness and the ability to generalize on unseen images. 
Note the ground truth contains more long texts than generations from the model since we set the max length to 64 during training and inference, which results in the discrepancy for text length greater than 60.

\Cref{fig:dist} (b) shows the keyword coverage (i.e.,~nouns, adjectives and adverbs) in output sentences. 
We consider an output as covering a keyword if the word exists in the corresponding ground truth.
We compare two models trained with and without $\mathit{PC^2L}$. 
We can see that $\mathit{PC^2L}$ improves the coverage of all kinds of keywords, which indicates our contrastive learning method diversifies and personalizes the generated text.
Overall, incorporating contrastive learning into multi-modal explanation generation leads to better output quality with more diverse and visually-aligned texts.




\xhdr{Can GPT-2 provide linguistic knowledge?}
Finally, we study whether GPT-2 can provide linguistic knowledge for our generation task. 
We train models with different weight initializations, with ground truth images (Img) or historical reviews (Text) as inputs. 
As shown in~\cref{tab:gpt-init}, comparing the performance of random and GPT-2 initialization, it is evident that the pretrained weights play a significant role. 
Finetuning on in-domain data (260k samples from users with one review and excluded from our personalization dataset) further improves domain-specific knowledge of the decoder and benefits generation performance on diversity metrics.

\begin{table}[]
\caption{Ablation Study on different initializations of the decoder. \textit{Random} randomly initializes model weights. \textit{Text GPT-2} and \textit{Img GPT-2} are initialized with weights from~\citep{Radford2019LanguageMA}. \textit{Img GPT-2 + FT} finetunes the model on a corpus similar to our training text data. Results are in percentage~(\%).}
\begin{tabular}{lccccc}
\toprule
\textbf{Method} & BLEU-1 & \textsc{Distinct-1} & \textsc{Distinct-2}  \\
\midrule
\textit{Img Random}    &    5.21    &  0.23   &   5.08   \\
\textit{Text GPT-2} &  4.81 &  3.43 & 19.27 \\
\textit{Img GPT-2}    &    \textbf{7.59}    &  4.05  &    29.41  \\
\textit{Img GPT-2 + FT} &  7.10    &  \textbf{4.32}  &   \textbf{30.82}  \\
\bottomrule
\end{tabular}
\label{tab:gpt-init}
\end{table}

\subsection{Case Study}
We study three examples (see~\cref{fig:case_study}) and compare our personalized showcases to single-modal explanations from \textit{Ref2Seq} and \textit{Text GPT-2}. Overall, our visual explanations is able to recommend images that fit users' interest. This indicates the effectiveness of our image selection module and the selected images can be used as valid visual explanations. 
More importantly, these images can provide grounding information for text generation such that the textual explanations become more informative (i.e.,~specific dishes), which aligns with our \textsc{CLIP-Align} metric as well as human evaluations in~\cref{sec:human}.
As is shown in~\cref{fig:case_study}, we can see historical review text alone 
cannot provide correct explanations (see Case 1) to the user (i.e.,~explanations from \textit{Ref2Seq} and \textit{Text GPT-2} are irrelevant to the user review) and the sentences are monotonous (see Case 2). In contrast, our showcase provides relevant and diverse textual explanations based on images. In case 3, our generated text missed some entities in the user's review since it only correctly describes one of the selected images. Hence, generating texts from multiple images is still a challenging problem for this new task.

As we can observe from the examples, \textit{Ref2Seq} tends to generate explanations with the same pattern, which also match the observation in~\cref{tab:task-2-main} that it has low \textsc{Distinct-1} and \textsc{Distinct-2}.

\begin{table}[t]
\caption{Human evaluation results on two models. We present the workers with reference text and images, and ask them to give scores from different aspects. Results are statistically significant via sign test~(p<0.01).}
\begin{tabular}{lcc}
\toprule
\textbf{Method} & Expressiveness & Visual Alignment  \\
\midrule
\textit{Ref2Seq}    &    3.72    &  3.65  \\
$\mathit{PC^2L}$   &   \textbf{4.25}    &  \textbf{4.10}    \\
\bottomrule
\end{tabular}
\label{tab:human}
\end{table}

\subsection{Human Evaluation}
\label{sec:human}
To fully evaluate our model, we conduct human evaluation on Amazon Mechanical Turk.\footnote{\url{https://www.mturk.com/}} For each model, we randomly sample 500 examples from the test set. Each example is scored by three human judges using a 5-point Likert scale to reduce variance. We instruct the annotators to consider two perspectives, expressiveness (semantically correct, diversity, no repetition) and visual alignment (the text describes the context of the images).
As is shown in~\cref{tab:human}, $\mathit{PC^2L}$ significantly outperforms \textit{Ref2Seq}, which is consistent with the automatic evaluation metrics. 



\section{Related Work}




\subsection{Explanation Generation}
There has been a line of work that studies how to generate explanations for recommendations~\citep{zhang2020explainable,yan2019cosrec}.
Some work generates product reviews based on categorical attributes~\citep{Zhou2017LearningTG} images~\citep{truong2019multimodal}, or aspects~\citep{Ni2018PersonalizedRG}. 
Due to noise in reviews, \citet{Li2019PersonaAwareTG} generated `tips' from the Yelp dataset which are more concise and informative as explanations in recommendation. 
To further improve the quality of generation, \citet{ni2019justifying} proposed to identify justifications by dividing reviews into text segments and classifying text segments to get ``good'' justifications. \citet{Li2021PersonalizedTF} proposed transformer-based model for recommendation explanation generations by incorporating user, item embeddings and related features.
These text generation tasks leverage historical reviews from users or items. 
Images, on the other hand, provide rich information and grounding for text generation. 
Moreover, multi-modal information in our task (i.e.,~images and text) are more acceptable than text as explanations for users. 

In this paper, we propose a new task for generating multi-modal explanations and present a framework that provides personalized image showcases and visually-aware text explanations for recommendations.

\subsection{Multi-Modal Learning}
Recent years have witnessed the success of deep learning on multi-modal learning and pretraining~\citep{radford2021learning,Chen2021VisualGPTDA}. These models usually adopt the Transformer~\citep{vaswani2017attention} structure to encode visual and textual features for pretraining, to later benefit the multimodal downstream tasks. Among them, CLIP~\citep{radford2021learning} is a powerful model trained on a massive amount of image-caption pairs, and has shown a strong zero-shot or transfer learning capability on various vision and language tasks, from image classification, image captioning, to phrase understanding~\citep{shen2021much,yan2022clip}. 
Several recent study~\citep{hessel2021clipscore,zhu2021imagine} used CLIP embeddings to compute modality similarities between image and text, and use CLIP-based scores as evaluation metrics for image captioning and open-ended text generation tasks. 

In our work, we used CLIP extensively as the multi-modal encoder for our framework. We also designed a new metric based on CLIP for evaluating the visual alignment between the image set and generated explanations. 


\subsection{Contrastive Learning}
The goal of contrastive learning~\citep{ oord2018representation} is to learn representations by contrasting positive and negative pairs. 
It has been investigated in several fields of applied machine learning, including computer vision~\citep{chen2020simple,he2020momentum}, natural language processing~\citep{huang2018large,gao2021simcse}, and recommender systems~\citep{xie2020contrastive,zhou2021contrastive,yan2022personalized}.
A few recent work showed promising results of applying contrastive learning to conditional text generation, by generating adversarial examples~\citep{lee2020contrastive}, finding hard negatives with pretrained language models~\citep{cai2020group,yan2021weakly}, or bridging image and text representations to augment text generation tasks~\citep{zhu2022visualize}.

Our work differs in that we study contrastive learning for conditional text generation in a cross-modal setting for personalization, where we proposed a novel contrastive framework for generating personalized multi-modal explanations.

\section{Conclusion}


In this paper, to generate explanations with rich information for recommendations, we introduce a new task, namely \emph{personalized showcases}, and collect a large-scale dataset \data from \emph{Google Local} for the task. We design a personalized cross-modal contrastive learning framework to learn visual and textual explanations from user reviews. Experimental results show that \emph{showcases} provide more informative and diverse explanations compared to previous text-only explanations. As future work, one promising direction is to develop an end-to-end framework for generating both visual and textual explanations. Besides, visual grounding on multiple images is still challenging for showcases. Another interesting setting is to address cold-start users or reviews written without images.
We hope our dataset and framework would benefit the community for future research on multi-modalities and recommendations.
\appendix
 
\section{Data Construction}
\label{app:data-collection}
Our dataset is constructed from \emph{Google Local} (i.e.,~maps) using a breadth-first-search algorithm with memorization. After collecting the review data, we filtered out reviews of length less than $5$ words, which are less likely to provide useful information; we also removed reviews ($2.13\%$) containing more than $10$ images. 
The details of \data-s1 construction for personalized image 
selection are as follows: We remove users with only one review for building a personalized dataset, then filter out reviews whose image urls are expired. After pre-processing, statistics for the personalized showcase dataset are shown in~\cref{tab:data}, where the number of images per business is 35.63 on average. We then randomly split the dataset by users, with 95,270/11,908/11,908 users for train/val/test.

\section{Visual Diversity Definition}
\label{sec:div-def}
We define the visual diversities in three levels as below:

\begin{itemize}
    \item \textbf{Intra-Business Div:} Measure the average diversity for image pairs at a business-level, where $\mathcal{P}_1(b)$ means all the possible image pairs for business $b$. $Z_1$ is the valid counts\footnote{When image set size is not more than 1, the dis-similarity calculation is invalid.} of dis-similarity calculations (same as below):
    \begin{equation}
       \sum_{b\in{B}}\sum_{m,n\in\mathcal{P}(b)}\frac{\textit{dis}(i^b_m, i^b_n)}{Z_1}.
    \end{equation}
    \item \textbf{Inter-User Div:} Measure the average diversity for image pairs from different users for the same business, where $\mathcal{P}_2(b)$ means all possible image pairs for business $b$ that come from different users:
    \begin{equation}
        \sum_{b\in{B}}\sum_{m,n\in\mathcal{P}_2(b)}\frac{\textit{dis}(i^b_m, i^b_n)}{Z_2}.
    \end{equation}
    \item \textbf{Intra-User Div}: Measure the average diversity in (business, user)-level, where $\mathcal{P}_3(u,b)$ means all possible image pairs from user $u$ to business $b$:
    \begin{equation}
        \sum_{b\in{B}}\sum_{u\in U}\sum_{m,n\in\mathcal{P}_3(u,b)}\frac{\textit{dis}(i^b_m, i^b_n)}{Z_3}.
    \end{equation}
\end{itemize}

\bibliographystyle{ACM-Reference-Format}
\bibliography{ref}


\end{document}